\documentclass[a4paper,twocolumn,showpacs,preprintnumbers,amsmath,amssymb,nofootinbib,floatfix]{revtex4}
\pdfoutput=1
\usepackage{hyperref}
\input psfig.sty
\usepackage{picinpar}
\usepackage{graphicx}

\begin{document}

\title{Reconstruction of cosmological matter perturbations in Modified Gravity}

\author{J. E. Gonzalez\footnote{E-mail: javierernesto@on.br}}

\affiliation{Observat\'orio Nacional, 20921-400, Rio de Janeiro - RJ, Brasil}

\date{\today}

\begin{abstract}
The analysis of perturbative quantities is a powerful tool to distinguish between different Dark Energy models and gravity theories degenerated at the background level.
In this work, we generalise the integral solution of the matter density contrast for General Relativity gravity \cite{SahniStarobinsky, Sahni2009} to a wide class of Modified Gravity (MG) theories.   To calculate this solution is necessary prior knowledge of the Hubble rate, the density parameter at the present epoch ($\Omega_{m0}$) and the functional form of the effective Newton's constant that characterises the gravity theory. We estimate in a model-independent way the Hubble expansion rate by applying a non-parametric reconstruction method to model-independent cosmic chronometer data  and high-$z$ quasar data.
In order to compare our generalised solution of the matter density contrast, using the non-parametric reconstruction of  $H(z)$ from observational data, with purely theoretical one, we choose a parameterisation of the Screened MG and the $\Omega_{m0}$ from WMAP-9 collaborations. Finally, we calculate the growth index for the analysed cases, finding very good agreement between theoretical values and the obtained ones using the approach presented in this work.

\end{abstract}


\maketitle

\section{Introduction}

The present cosmic  accelerated phase of the Universe discovered at the end of the 90's \cite{Riess,Perlmutter} can be explained by the standard cosmological model, the so-called $\Lambda$CDM. In this model, the content of the Universe is mainly composed  by  Dark Energy (DE)   characterised by  the cosmological constant ($\sim 70\%$), dark matter ($\sim 25\%$) and ordinary matter ($\sim 5\%$) \cite{review}.
However, there are some inconsistencies in the standard model from the theoretical point of view (e.g. fine tuning and cosmic coincidence problems \cite{Weinberg}) and some tensions in the observational 
constraints (see Ref \cite{Lin} for a brief review). For these reasons, many models describing the behaviour of dynamical DE \cite{Frieman2008,BAZR,BA,CPL-Polarski,CPL-Linder,Jassal} and  modified gravity (MG) (see Ref. \cite{Capozziello, Clifton:2011jh} for reviews) theories at large scales have been proposed to explain satisfactorily the cosmic acceleration.
The nature of these two approaches is essentially different. The first one constitutes a modification of the r.h.s. of the Einstein Equations (EE) where it is included a highly negative pressure fluid to the Universe content. 
The second one corresponds to a modification of the geometrical side of the EE. In both of them, the DE or the MG can be described by the inclusion of  scalar fields (e.g chameleons \cite{chameleon}, symetrons \cite{symetron}, dilatons \cite{dilaton}, galileons \cite{galileon}). The effect of these fields is tightly constrained by local experiments, which implies that they are screened in dense environments \cite{Brax1,Brax2}.

The accuracy and the number of geometrical cosmological observables have grown considerably.
Nevertheless, this is not enough to distinguish between some DE and MG theories because they can predict the same cosmic expansion and the same values for background observables like the  luminosity and angular diameter distance. Hence, it is necessary to explore some quantities at the perturbation level where the background degeneracy between different explanations of the cosmic acceleration may be broken \cite{Wang2008}. For instance, in the screened MG theories the cosmic expansion is the same as the one in the $\Lambda$CDM but the  perturbative quantities growth presents an anomalous behaviour inside the Compton radius of the scalar field \cite{Brax2}.

Each DE model or MG theory assumes a specific functional behaviour of dynamical and kinematic variables, being the parameters of the theory constrained by observational data. In contrast, it is possible to obtain model-independent information directly from the observational data using non-parametric methods \cite{NP,Rasmussen,Holsclaw2010,Holsclaw2011, Seikel,Seikel2012,BustiH0,Shafieloo2006,Shafieloo2007,Shafieloo2012a,Li,Gonzalez1,Gonzalez2,Nisha}. In the latter scenario, it is assumed a correlation between each data point but it is not required prior information about the functional form of the observable. 

In this work, we generalise the integral solution of the linear matter density contrast valid for  General Relativity gravity in a homogeneous and isotropic universe, presented in Ref. \cite{SahniStarobinsky, Sahni2009},  for  MG theories where the matter decays proportional to $a ^ {-3}$ and also the effect of the MG is encoded in the effective Newton's constant, $G_{eff}(k,a)$. This solution requires the knowledge of the Hubble parameter, the matter density contrast at the present epoch and the functional form of the MG effect. In order to obtain a model-independent reconstruction of the Hubble parameter, we apply the NPS method  \cite{Shafieloo2006,Shafieloo2007,Shafieloo2012a,Li,Gonzalez1,Gonzalez2,Nisha} to the cosmic chronometer data at the redshift range [0.07, 1.04] \cite{Jimenez03,Simon05,Stern10,Moresco12,Moresco:2015cya,Moresco2016} and the high-$z$ quasar data at $z\approx 2.3$ \cite{Busca}. The cosmic chronometer data, obtained from the differential age method for passively evolving galaxies, within its redshift range is aimed to be model-independent both cosmological and stellar population synthesis. The quasar data is obtained by the correlation function of the transmitted flux fraction in the  Ly$\alpha$-forest of high-$z$ quasars.  We combine these two data sets to obtain information of the behaviour of $H(z)$ up to the matter dominated epoch.

We reconstruct the matter perturbation quantities following the approach presented in Ref. \cite{Sahni2009,Gonzalez1,Gonzalez2}  for the same MG parameter values used in Ref. \cite{Brax2}. For this, we assume the model-independent reconstruction of the cosmic rate and  the parameterisation of screened MG proposed in Ref. \cite{Brax1,Brax2}. This approach   
 allows us to  explore the validity of this screened MG using only background data, by comparing the purely theoretical calculations and the ones obtained in this work. 

This paper is organised as follows: in Sec \ref{MPE} we generalised the treatment of linear matter perturbations of Ref. \cite{SahniStarobinsky,Sahni2009}
and present the considered screened MG  to reconstruct the perturbative quantities.  We also introduce the basic equations of the matter perturbation theory. In Sec. \ref{DaHPR} we discuss the observational data and present the non-parametric method used to reconstruct the cosmic expansion history.
In Sec. \ref{R} we present the results of our reconstructed cosmic expansion and the matter perturbation analysis. We end this paper with the main conclusions in Sec. \ref{C}.

\section{Matter Perturbation Equations}
\label{MPE}

In order to study the behaviour of the matter perturbations in MG theories at large scales, we consider the parameterisation \{$m(a), \beta(a)$\} of screened MG presented in \cite{Brax1,Brax2}. 
These models are defined by the effect on the evolutions of the matter perturbations, whereas the dynamics of the Universe is the same as in the $\Lambda$CDM model at the background level. In this approach, the gravity is modified 
on large scales by a scalar field and it remains  unchanged in dense environments.

In Ref \cite{Brax2}, the authors obtain the  equations that govern the evolution of the matter perturbations  in  a   homogeneous and isotropic universe with the effects of a screened MG. In this case,   the scalar  perturbations can be described by the metric

\begin{equation}
 ds^2=-(1+2\Psi)dt^2+(1-2 \Phi)a^2(t)d \vec{x} ^2\;,
\end{equation}
where  $\Phi$ and $\Psi$ are the scalar potentials in the longitudinal gauge. Also, we  assume that the potentials are proportional by a scale and time dependent 
function $\Gamma(k,a)$ as:

\begin{equation}
\Psi_k = \Gamma(k,a) \Phi_k.
\end{equation}
where the subindex $k$ represents the Fourier-space quantity.

In this scenario, the Poisson equation is modified by the parametric function $\mu(k,a)$ as:

\begin{equation}
-k^2 \Psi_k= 4\pi \mu(k,a)G_N \rho_m \delta_k, 
\end{equation}
where $G_N$, $\rho_m$ and $\delta_k $ correspond to the Newton's constant, the background matter density and the density contrast ($\delta(\vec{x},t)\equiv (\rho(\vec{x},t)-\rho(t))/\rho(t)$), respectively.

Finally, it is found  the second order differential equation for matter perturbations \cite{Brax2} \footnote{In the rest of the text we ignore the Fourier space subindex $_k$}

\begin{equation}
\label{2odeA}
\delta''+\frac{a'}{a}\delta'- \frac{3\Omega_{m}(a)}{2}\left(\frac{a'}{a}\right)^2 \mu(k,a) \delta = 0\;,
\end{equation}
or 

\begin{equation}
\label{2odeA2}
\ddot{\delta}+2H\dot{\delta}-\frac{3\Omega_{m}(a)}{2}H^2 \mu(k,a) \delta = 0\;,
\end{equation}
the prime denotes the derivative with respect to the conformal time and the dot the derivative with respect to the cosmic time. The $\Omega_{m}(a)$ is the matter density parameter defined by:
\begin{equation}
 \Omega_m(a) \equiv \frac{\Omega_{m0}H_0^2}{a^3 H^2(a)} \;,
 \label{omdef}
\end{equation}
and $H$ is the Hubble parameter. In this case, the evolution of the matter structures is totally determined by the functions $H$ and $\mu(k,a)$, and the value of the matter density today $\Omega_{m0}$. In these equations, any effect of the MG  is encoded in the function  $\mu(k,a)$.

Defining the dimensionless physical distance,

\begin{equation}
D=H_0 \int_t^{t_0}\frac{dt}{a(t)}=H_0\int_0^z \frac{dz_1}{H(z_1)}\;,
\end{equation}
we can rewrite the Eqs. (\ref{2odeA}-\ref{2odeA2}) in terms of the redshift and $D$ as:

\begin{equation}
\frac{d}{dD}\left(\frac{d \delta/ dD }{1+z(D)}\right)=\frac{3}{2}\Omega_{m0}\mu(k,a)\delta,
\end{equation}
assuming that $\rho_m \propto a^{-3}$.

The solution of the previous equation can be written as an integral function as:

\begin{subequations}
\begin{eqnarray}
 \label{22odes}
\delta(D)&=&1+\delta_0 '\int_0 ^D[1+z(D_1)]dD_1  +\frac{3}{2}  \Omega_{m0}\\ &&
\times \int_0^D[1+z(D_1)]\left(\int_0^{D_1}\mu(k,a)\delta(D_2)dD_2  \right)dD_1 \nonumber \;,
\end{eqnarray}
\begin{eqnarray}
\delta'(D)&=&\delta_0 '[1+z(D)] \\ && +\frac{3}{2}\Omega_{m0}[1+z(D)] 
\int_0^{D}\mu(k,a)\delta(D_1)dD_1 \nonumber \;.
\end{eqnarray} 
\end{subequations}
which are easily calculated if we know the Hubble parameter and $\mu(k,a(D))$.

The particular form of the function $\mu(k,a)$ depends on the values of the parameters \{$m(a), \beta(a)$\}. In this parameterisation, $m(a)$ represents the mass of the scalar field at the background level and $\beta(a)$ represents the coupling function between the field and the CDM particles \cite{Brax2}. In this paper, we consider the same family of parameterisation than Ref. \cite{Brax1, Brax2}, where the screened MG is characterised by

\begin{equation}
 \mu(k,a)=\frac{(1+2\beta(a)^2)k^2+m(a)^2a^2}{k^2+m(a)^2a^2}\;,
 \label{mu}
\end{equation}
with $\beta= 1/\sqrt{6}$ and the mass of the field given by:

\begin{equation}
 m(a)= m_0a^{-3(n+2)/2},
\end{equation}
where $m_0$ is a free scale close to 1 Mpc$ ^{-1}$ and $n>1$.

Now, if we identify 

\begin{equation}
 \frac{G_{eff}(k,a)}{G_N}=\mu(k,a),
\end{equation}
the Eq (\ref{22odes}) constitutes a generalization of the solution for the density contrast shown in Ref \cite{Sahni2009} for any MG theory satisfying the second order differential equation
\begin{equation}
\label{2odeB}
\ddot{\delta}+2H\dot{\delta}-4 \pi G_{eff} \rho_m = 0\;,
\end{equation}
where $G_{eff}$ is the effective Newton's constant.

Thus, we define the growth rate  as:

\begin{equation}
 f(k,z)\equiv\frac{d\ln \delta(k,z)}{d\ln a}=-\frac{(1+z)H_0}{H(z)}\frac{\delta'(k,z)}{\delta(k,z)}\;,
 \label{fdef}
\end{equation}
where $z$ is the redshift.
This quantity has crucial importance because it is directly estimated by the observations of the large-scale structure (LSS) of the Universe. 
Due to the growth rate constitutes a cosmological observable, it is possible to compare the one obtained from solving the previous equation using background information
with the one obtained from the LSS observations.
This comparison allows us to prove the hypothesis of the background dynamics of the Universe used to calculate the Eq. (\ref{2odeA}). These hypotheses are:

\begin{itemize}
 \item The cosmological principle is valid, therefore, the geometry of the Universe at the background level is described by the unperturbed FLRW metric. For example, in inhomogeneous geometries the equations are modified as presented in Refs. \cite{Marteens}
 
 \item The matter energy-momentum tensor is covariantly conserved. This implies that the matter density decays proportional to $a^{-3}$. In models in which there is interaction in the dark sector, the dark matter does not decay at the same rate and, in the general case, the differential equation have to be modified \cite{Mehrabi2,Rsousa,Nesseris2017b}. 
 
 \item The correct theory of gravity is the Screened MG characterised by the function $\mu(k,a)$  at the perturbation level (Eq. (\ref{mu})) and behaved as a $\Lambda$CDM cosmology at the background level. As mentioned below, we can generalise this condition to any MG theory satisfying the Eq. (\ref{2odeB}).
 
\end{itemize}

Finally, we complete the set of matter perturbation quantities defining the growth index:
\cite{Peebles,Lahav,Wang1998}
\begin{equation}
\gamma(k,z) = \frac{\ln{f(k,z)}}{\ln{\Omega_{m}(z)}}\;,
 \label{gammadef}
\end{equation}
which is an excellent tool to characterise modified gravity theories. For instance, we highlight the growth index value for the following models:  $\Lambda$CDM model, $\gamma_0=6/11$; slow varying $w$CDM DE models, $\gamma \simeq \frac{3(w-1)}{6w-5}$ \cite{Wang1998,LinderCahn} and  DGP model, $\gamma_0=11/16$ \cite{LinderCahn}. In these cases, the growth index is well described by a constant function \cite{Polarski2016}. However, in a general way, it can evolve with the redshift and for future accurate data its evolution must be taken into account \cite{Wang2008}.

\section{Data and Hubble Parameter Reconstruction}
\label{DaHPR}

\subsection{Data}
\label{D}

In the approach presented in this work to calculate the cosmological matter perturbations, the expansion history of  the Universe plays a crucial role.
Therefore, it is very important to use a model-independent data to obtain the cosmic expansion rate.  For this purpose, we use the cosmic chronometers estimates  based in the differential age inferences \cite{Jimenez-Loeb}. 

The basic assumption of this method is that  the relative age of two passively-evolving elliptical red galaxies at approximately the same redshift can be used to estimate the age variation of the Universe and, consequently,  the Hubble expansion rate at the redshift of the observation. The difference between the two galaxies ages corresponds to the variation of the  age of the Universe, $\Delta t$, and with the redshift difference, $\Delta z$, it is possible to approximate the Hubble parameter as: $H(z) \simeq -1/(1+z)\Delta z/\Delta t$. The differential age approach estimates the Hubble rate directly from the data without assuming a specific spatial geometry or any other cosmological model.

The cosmic chronometers estimates are cosmological model-independent, but  they can  depend on the stellar population synthesis models at high redshift. As pointed out in Ref \cite{Licia}, the Hubble parameter measurements are almost stellar population synthesis model-independent until $z\simeq1.2$. Following the discussion in Ref \cite{Licia}, we use the cosmic chronometer data up to $z = 1.04$ \cite{Simon05,Jimenez03,Stern10,Moresco12,Moresco:2015cya} and we increase (20\%) the error bar of the data point at $z=1.04$. Note that we also use the cosmic chronometer data $H(z=0.4293)=91.8\pm 5.3$ \cite{Moresco2016}. 

\begin{figure*}[t]
\includegraphics[width = 8.7cm, height = 6.5cm]{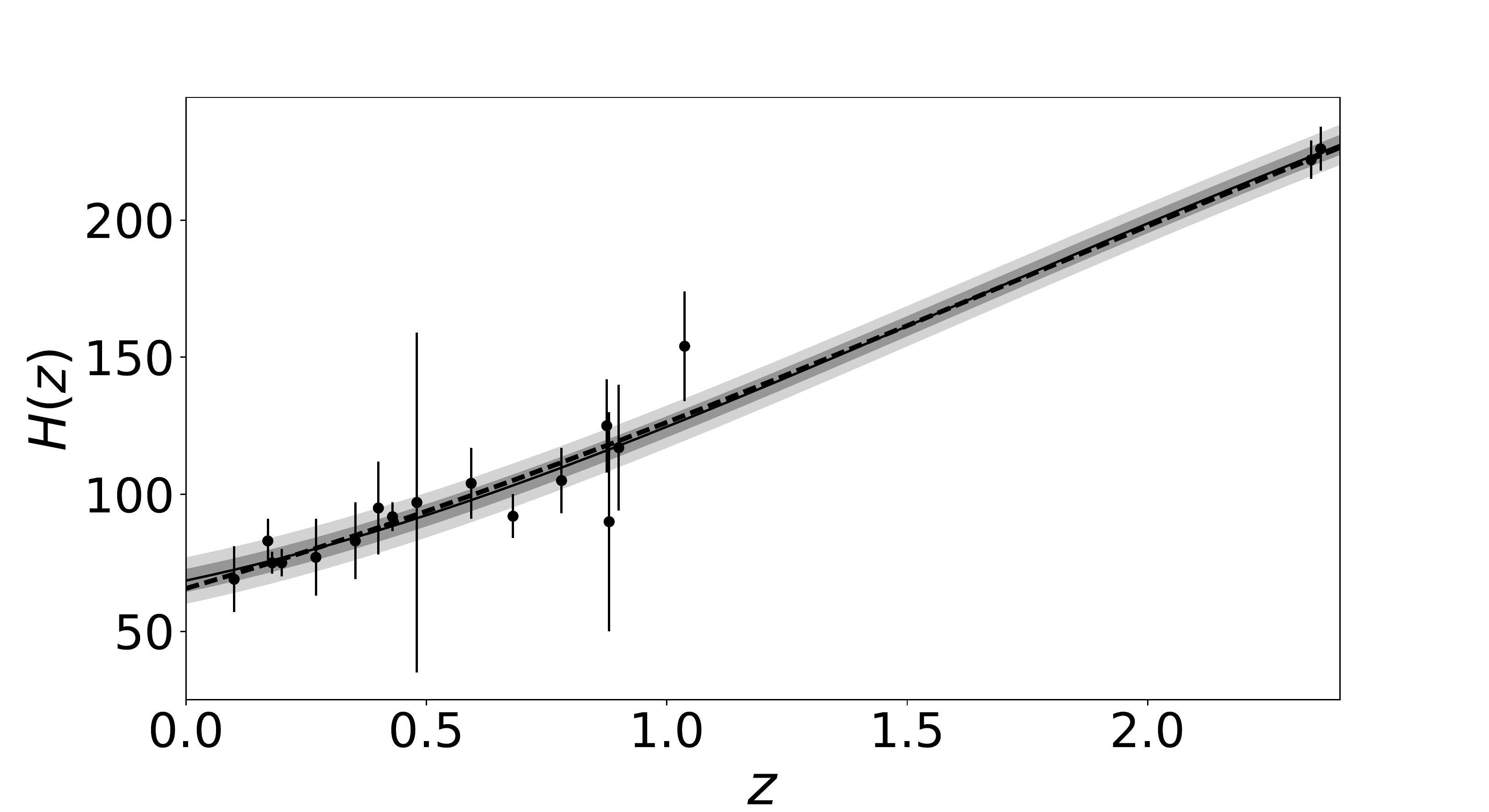}
\hspace{0.1cm}
 \includegraphics[width = 8.7cm, height = 6.5cm]{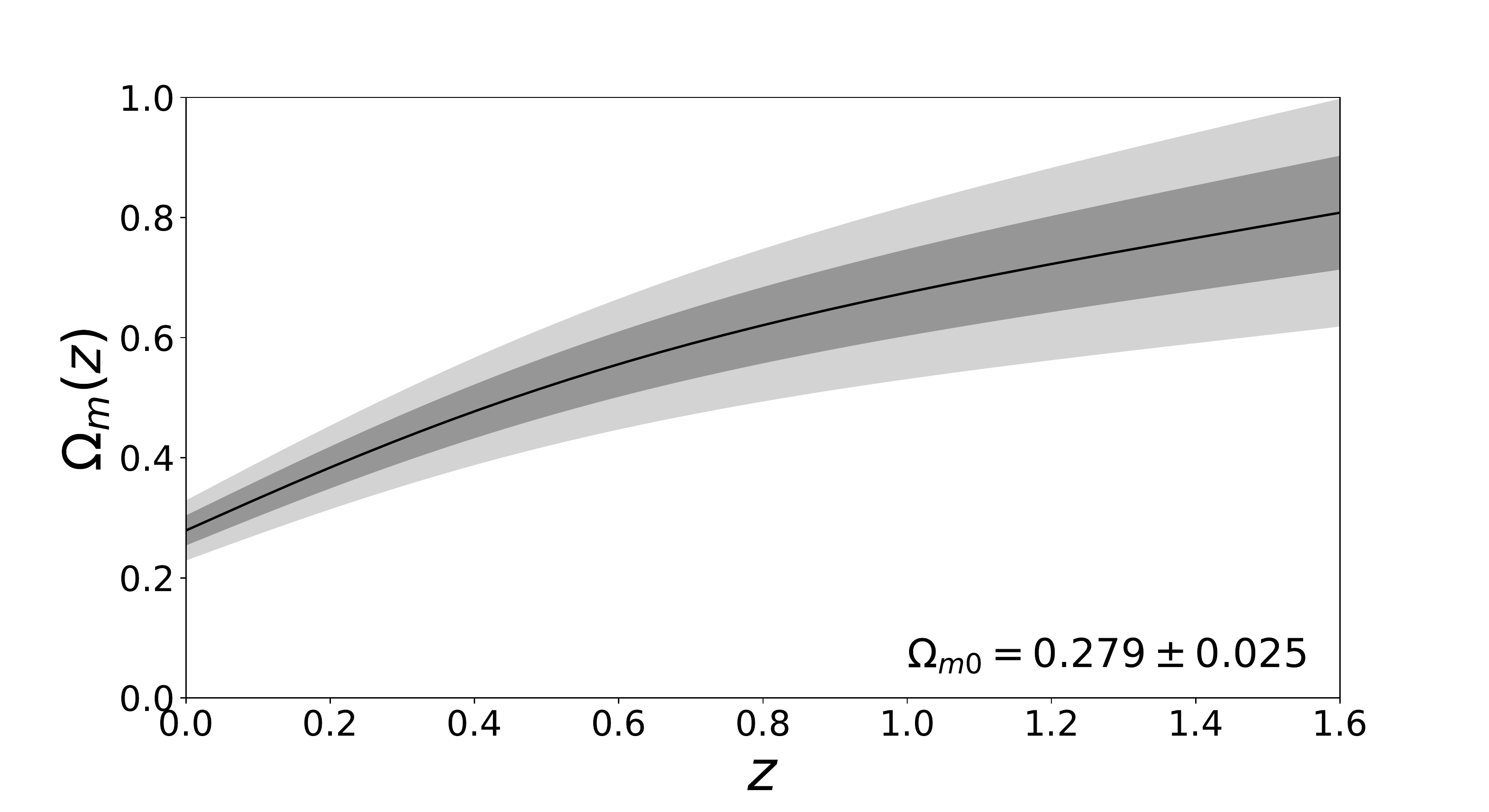}
\caption{a) The reconstruction of the Hubble parameter using the cosmic chronometer and high-$z$ quasar data. The solid line   correspond to the reconstruction applying the NPS method (Sec. \ref{NPS}) and the dashed line correspond to a reconstruction using GP method. b) The matter density parameter using the NPS reconstruction of $H(z)$ and the $\Omega_{m0}$ prior from WMAP-9 collaborations \cite{WMAP9}. The shaded regions represent the $1\sigma$ and $2\sigma$ confidence levels. }
\label{HZ}
\end{figure*}

Finally, we complete the dataset with two measurements of the Hubble parameter at high-$z$, $z = 2.34$ \cite{Debulac} and $z = 2.36$ \cite{Font-Ribera}, obtained from the correlation function of Ly$\alpha$-forest systems from quasar data. The complete data set is shown in Fig. \ref{HZ}.

\begin{figure*}[t]
\includegraphics[width = 8.7cm, height = 6.5cm]{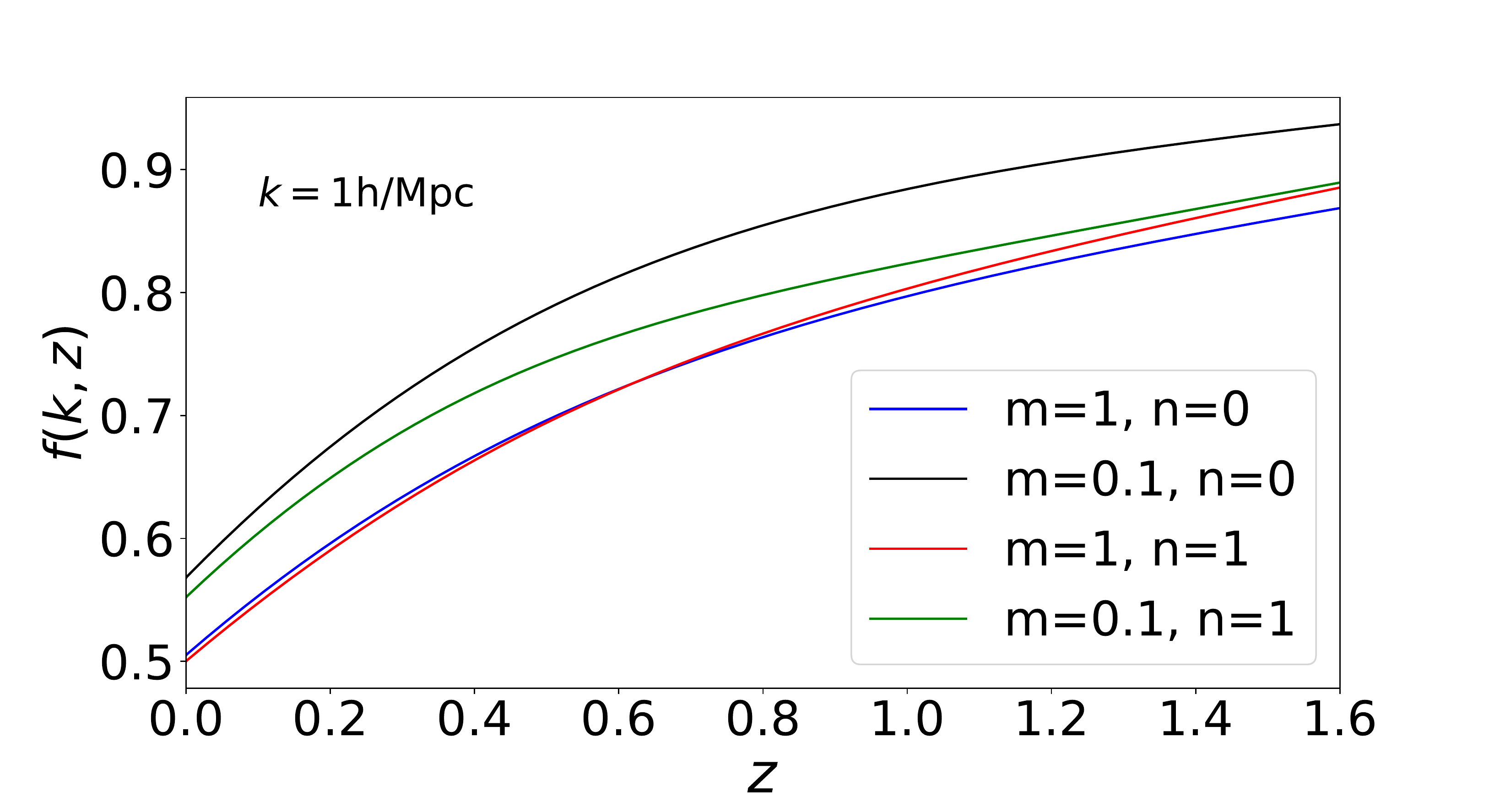}
\hspace{0.1cm}
\includegraphics[width = 8.7cm, height = 6.5cm]{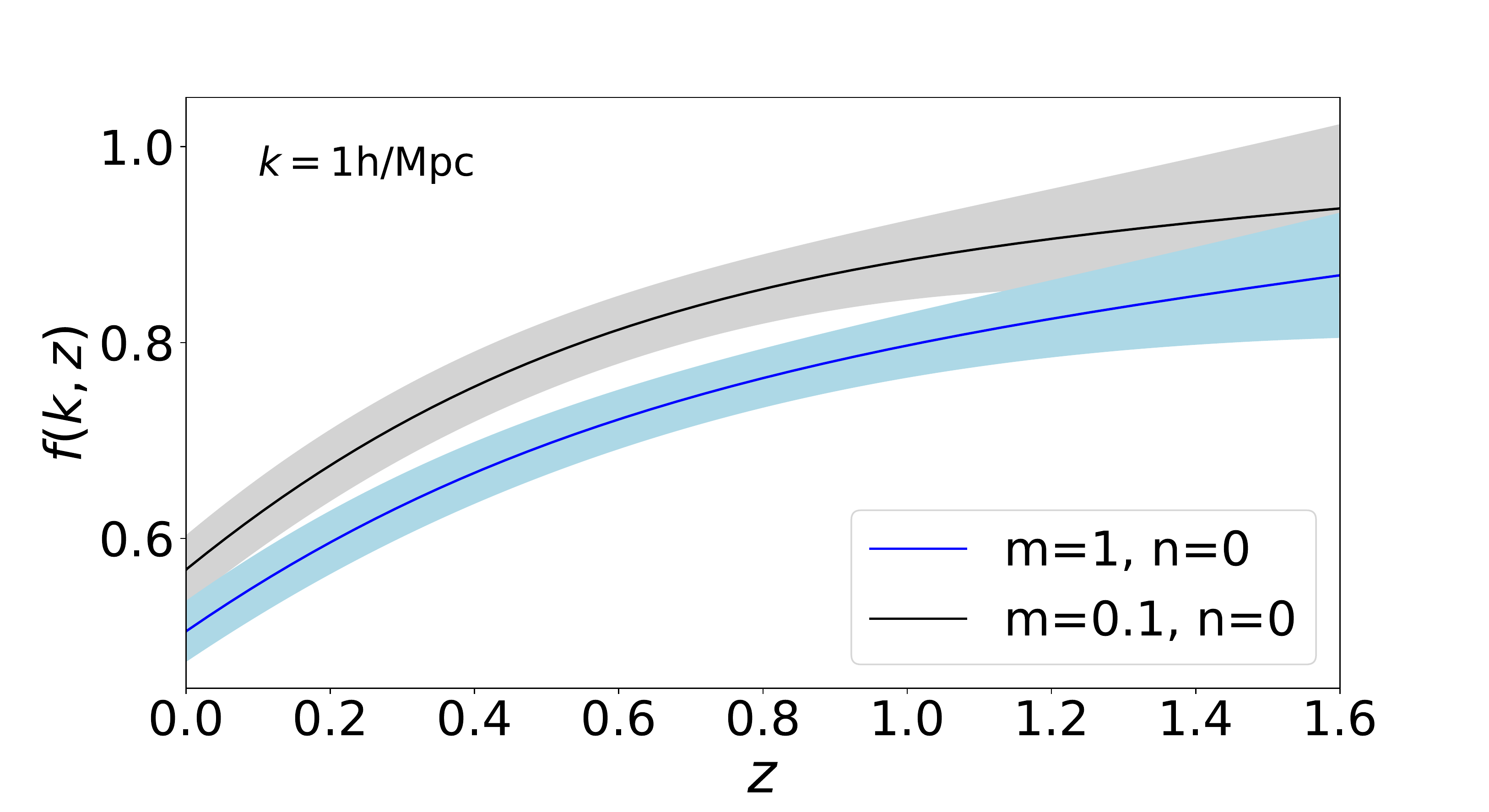}
\caption{a) The growth factor on sub-Hubble scale obtained solving Eq. \ref{22odes} using the WMAP-9 $\Omega_{m0}$ value and the parameterisation of the screened MG in Eq. (\ref{mu}). b)The same as in the previous panel for just two values of ($m_0, n$).  The solid line corresponds to the reconstruction whereas the shaded regions  represent the 1$\sigma$ confidence level. The unit of $m_0$ is Mpc$^{-1}$.}
\label{fz}
\end{figure*}

\subsection{Non-parametric Smoothing}
\label{NPS}
To obtain a smooth function in a non-parametric way that represents the expansion rate of the Universe until very high redshift, we apply the  NPS method, firstly proposed in Ref \cite{Shafieloo2006}, to the cosmic chronometer and high-$z$ quasar data. This method has been widely used in the literature to reconstruct
the luminosity distance, the physical distance and the Hubble parameter.

The main goal of this NPS method is to smooth the noise of the data. For this, we subtract an initial guess model to the data and then apply the smoothing kernel. We recover the quantity adding back the guess model. The general form of the method taking into account the data errors was presented in Ref \cite{Shafieloo2012a}. The smoothing function is obtained with the expression:
\begin{equation}
 H^s(z,\Delta) = H^g(z) + N(z) \sum_i \frac{\left[H(z_i)- H^g(z_i) \right]}{\sigma^2_{H(z_i)}} \times {\cal{K}}(z,z_i),
\end{equation}

\noindent where $H^s(z,\Delta)$ is the reconstructed smoothed function,  $H^g(z_i)$ is the 
 guess model, $H(z_i)$ is the observational data, $\sigma_H(z_i)$ is the 
data error, $\Delta $ is the smoothing scale and $N(z)$ is the 
normalization factor given by:
\begin{equation}
 N(z)^{-1}= \sum_i \frac{{\cal{K}}(z,z_i)}{\sigma^2_{H(z_i)}}.
\end{equation}

\noindent As made in Refs \cite{Li, Gonzalez2}, we adopt a Gaussian kernel (${\cal{K}}(z, z_i)=\exp(-(z-z_i)^2/2\Delta^2)$) to perform the reconstruction.

Due to the fact the  selection of the initial guess model is arbitrary, we must apply the NPS method repeatedly modifying in each iteration the guess model for the function obtained in the previous step. 

We can consider that for any reliable initial guess the reconstruction converges to the same function \cite{Shafieloo2006,Shafieloo2007,Li,Nisha}.

In order to obtain the optimal value of the smoothing scale, we calculate the cross validation function,
\begin{equation}
 CV(\Delta)=\frac{1}{n}\sum_i (H(z_i)-H^s_{-i}(z_i|\Delta))^2,
\label{CV}
\end{equation}
and we choose the $\Delta$ which minimizes it. This scale controls how smooth is our reconstruction. Finally, to calculate the error, we follows the approach in Refs \cite{Li, Gonzalez2}

\begin{figure*}[t]
\includegraphics[width = 8.8cm, height = 6.3cm]{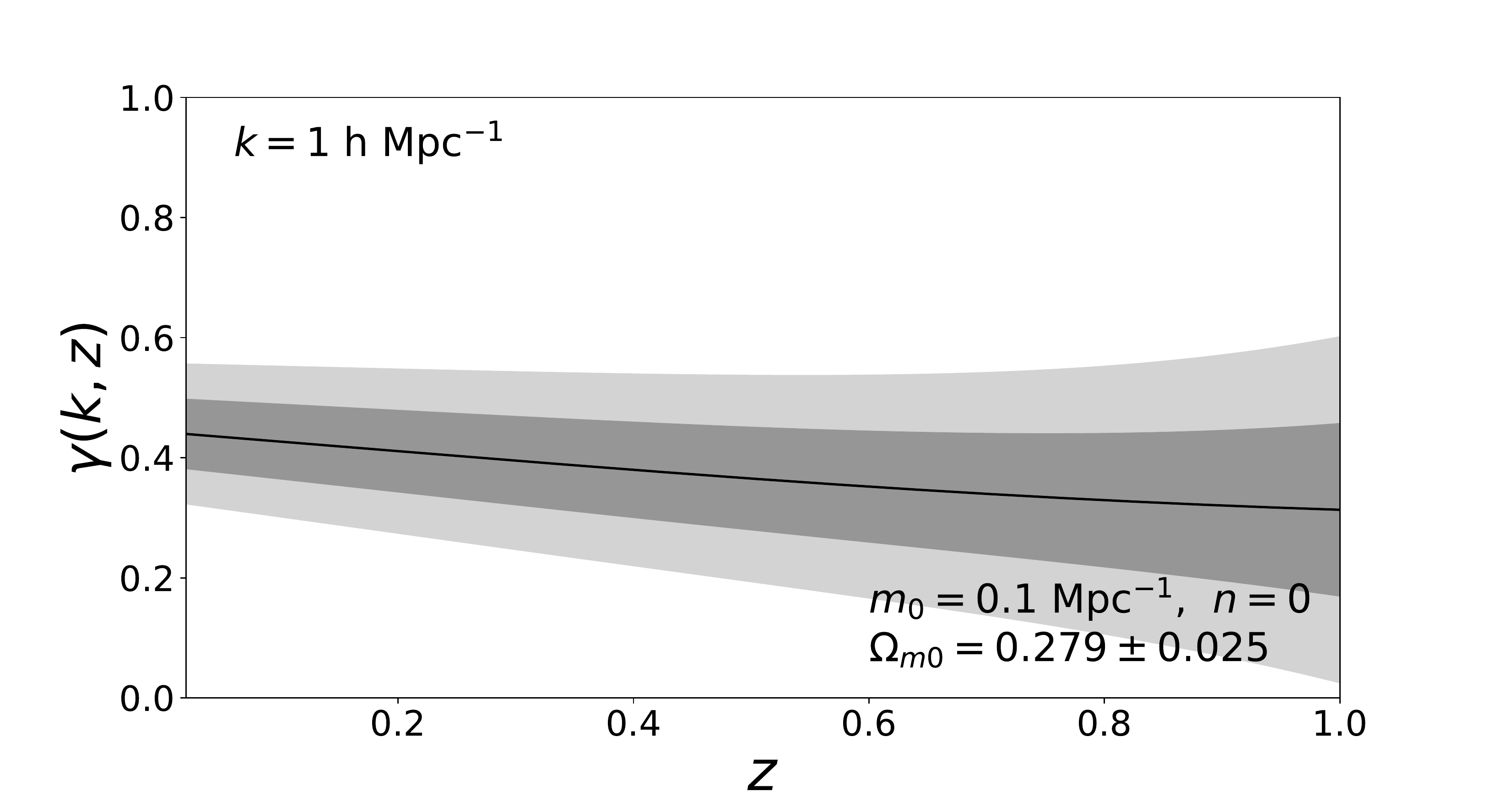}
\hspace{0.1cm}
\includegraphics[width = 8.8cm, height = 6.3cm]{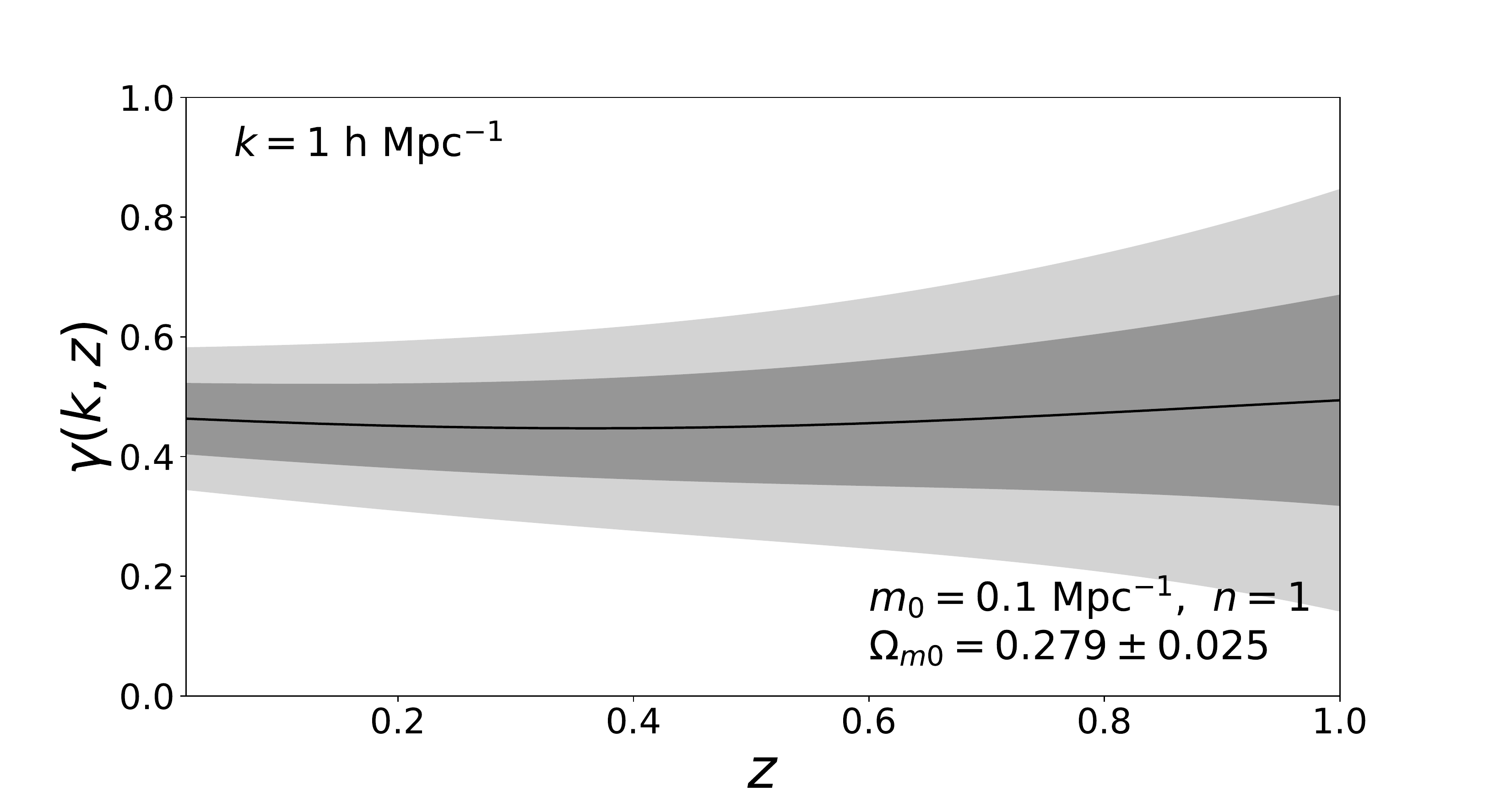}
\hspace{0.1cm}
\includegraphics[width = 8.8cm, height = 6.3cm]{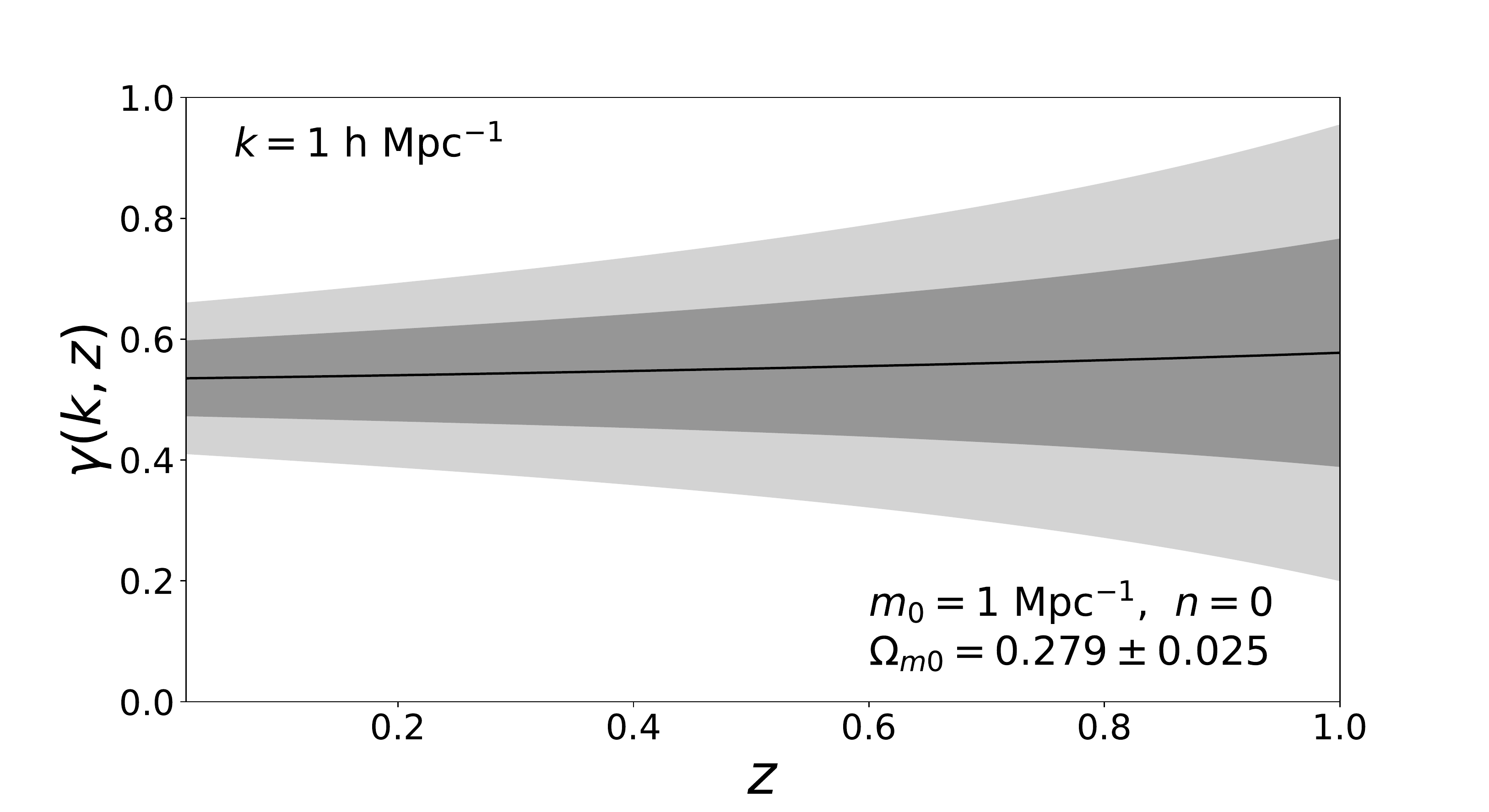}
\hspace{0.1cm}
\includegraphics[width = 8.8cm, height = 6.3cm]{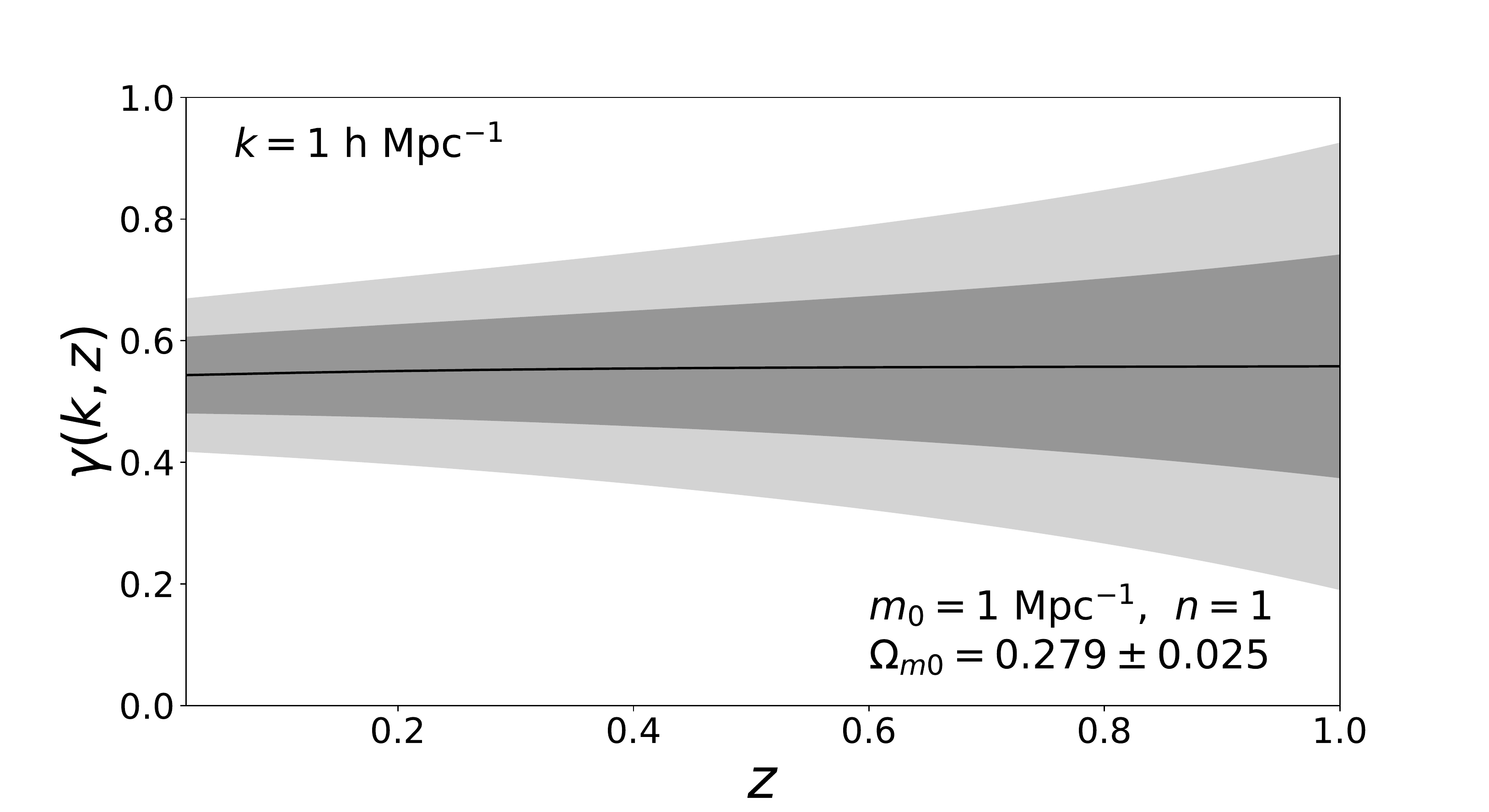}
\caption{ The growth index calculated solving Eq. \ref{22odes}  for four combinations of the parameters $m_0$ and $n$ using the WMAP-9 $\Omega_{m0}$ value. The solid line corresponds to the reconstruction whereas the shaded regions  represent the 1$\sigma$ and 2$\sigma$ confidence levels. }
\label{gamma}
\end{figure*}

\section{Results}
\label{R}

We apply the NPS method described in Sec. \ref{NPS} to the data in Sec. \ref{D} obtaining a smooth function for the cosmic expansion rate. As the reconstruction method and the data are model-independent, it is expected that our estimate of the Hubble parameter is not biased by any cosmological model. In Fig. \ref{HZ}, we show the final result of the NPS method and, for comparison, we also plot the result obtained using the non-parametric method Gaussian Process (GP) \cite{Rasmussen,Holsclaw2010,Holsclaw2011,Seikel2012,Gonzalez1}. As it was shown in Ref \cite{Gonzalez2}, the NPS reconstruction of these data presents a decelerated phase at high redshift which is expected in the standard cosmology when the matter domains the Universe dynamics. We emphasize that this behaviour is required  because the assumptions used to solve the differential equation of the density contrast. We also present the matter density parameter calculated with the Eq.(\ref{omdef}) \cite{Gonzalez2} using the reconstruction of $H(z)$ and the value of the density parameter at the present epoch from WMAP-9 collaborations \cite{WMAP9}. 

Now,  to solve the Eq.(\ref{22odes}), we need to choose a value for $\Omega_{m0}$ and a parameterisation for the MG. We employ the parameterisation shown in Eq.(\ref{mu}). In order to perform a directly comparison of our results from a non-parametric reconstruction of the Hubble parameter and the theoretical prediction of the matter perturbations in this screened MG scenario, we explore the same values as in Ref \cite{Brax2} for ($m_0,n$)=(1 Mpc$^{-1}$,0),(0.1 Mpc$^{-1}$,0),(1 Mpc$^{-1}$,1), (0.1 Mpc$^{-1}$,1) and the same $\Omega_{m0}$ value from WMAP-9 \cite{WMAP9}. With a fix gravity theory and matter density prior, we solve the Eq.(\ref{22odes}) for many values of $\delta_0'$ and select the one that produces $(1+z)\delta \propto cons$ at the redshift of the last $H(z)$ observation (see Refs. \cite{Gonzalez1,Gonzalez2} to know more details). This condition is required by the standard theory of cosmological perturbations in a matter domination epoch and compatible with the considered screened MG.

We obtain the growth rate for four values of the MG parameters using the calculations of the density contrast. These results are shown in Fig. \ref{fz}a). We find a good agreement between the results herein obtained and the theoretical ones in the Fig. 5 of the Ref. \cite{Brax2}. Nevertheless, we recognize a difference in  the interception of the curves with ($m_0,n$)=\{($1,0$), ($1,1$)\} and their behaviours at high-$z$.  In Fig. \ref{fz}b), we plot the growth rate for the MG parameters ($m_0,n$)=($1,0$), ($0.1,0$) with their respectively 1$\sigma$ confidence  level follows the approach proposed in this work. It shows that the Hubble parameter information can distinguish between these two MG parameters at $1\sigma$ level. The comparison of these results with the observational data of the growth rate can constrain the parameters of the  considered gravity theory or, in a more general way, other MG theories.

 \begin{table}
	\begin{center}
 		\begin{tabular}{lccccc}
 		\hline\hline
$m_0$& $n$ & & $\gamma_0^{teo}$ & &$\gamma_0^{rec}$ \\
 \hline
 1 Mpc$^{-1}$& 1 & & 0.54 & & 0.54 $\pm$ 0.06 \\
 0.1 Mpc$^{-1}$& 1 & & 0.47 & & 0.46 $\pm$ 0.06 \\
  1 Mpc$^{-1}$ &0 & & 0.54 & & 0.53 $\pm$ 0.06 \\
  0.1 Mpc$^{-1}$& 0 & & 0.44 & & 0.44$\pm$ 0.06\\
 \hline\hline
 		\end{tabular}
 	\end{center}
 	\caption{Calculations of the growth index via the reconstruction of the matter density perturbation and the theoretical value by \cite{Brax1}.}
 	\label{gammat}
 \end{table}

Finally, we calculate the growth index (Eq.(\ref{gammadef})) and plot it in Fig. \ref{gamma}. Note that the growth index for the parameters ($m_0,n$)=\{($1,0$), ($1,1$)\} does not evidence  redshift (time) evolution and in this case $f \approx \Omega_m(z)^{\gamma_0}$ would be a good approximation of the growth rate. In contrast, for the screened MG parameters ($m_0,n$)=\{($0.1,0$), ($0.1,1$)\}, there is evolution of the growth index which is compatible with what is expected in theories beyond GR \cite{GannoujiPolarski,Motohashi}. As pointed out in Ref. \cite{PolarskiGannouji},  the derivative of the growth index can be also used to characterised gravity theories. In Table \ref{gammat}, we present the values of $\gamma_0$ for the analytical approach \cite{Brax2} and the ones obtained with the calculations of the matter density perturbations via non-parametric reconstruction of the expansion rate.

\section{Conclusions}
\label{C}

In this work we have generalised  the integral solution of the matter density contrast at the linear regime. This generalisation is valid for a MG theory that satisfies the Eq.(\ref{2odeB}) and that conserves a matter dominated epoch in the past of the Universe history.

We have performed a non-parametric reconstruction of the Hubble parameter. To perform this reconstruction, we have applied the NPS method (Sec. \ref{NPS}) to cosmic chronometers and high-$z$ quasar data (Sec. \ref{D}) at the redshift range [0.07, 2.36]. By analysing the data we obtain information about the cosmic rate until the redshift where it is expected that the matter domains according the $\Lambda$CDM model.  We have calculated the matter perturbation quantities using the $H(z)$ model-independent reconstruction, the $\Omega_{m0}$ value from WMAP-9 and assuming the $\mu(k,a)$ parameterisation of MG. Our results of the growth rate (Fig. \ref{fz}) have shown a very good agreement with the theoretical ones presented in Ref. \cite{Brax2}. These results evidence the validity of the generalised solution of this work.  It is expected that our results are minimally biased due to the fact that the reconstruction method and the data are model-independent.

As shown in Fig \ref{fz}b) different values of the parameters of $\mu(k,a)$ can be distinguished by $H(z)$ data  at 1$\sigma$ confidence level analysing the growth rate. We will be able to test the validity of the fundamental hypotheses envolved in the analysis (Sec. \ref{MPE}) by comparing the matter perturbation reconstruction with future scale dependent growth rate estimates.

We have calculated the growth index and plotted it in Fig. \ref{gamma}. We have found that for the cases ($m_0,n$)=\{($1,0$), ($1,1$)\} the growth index is degenerated and it can be well described by a constant. The other two considered cases are univocally characterised and revealing a possible evolution of the growth index. The calculated values of the growth index at the present epoch are compatible with the theoretical ones (see Table. \ref{gammat}).
Just for comparison, we calculate the growth index using a non-parametric reconstruction of $H(z)$ via GP method. We found that the differences in the growth indexes are not higher than 5\%. This exposes the robustness of the methods applied.

\begin{acknowledgments}
JEG thanks CAPES and FAPERJ (Brazilian agencies) for the grants under which this work was carried out. JEG also thanks J. S. Alcaniz, R. S. Gon\c calves and X. Saad-Olivera for the useful discussions.
\end{acknowledgments}

\end{document}